# Broadband Tunable Deep-UV Emission from AI-Optimized Nonlinear Metasurface Architectures

Omar A. M. Abdelraouf [a,*]

[a] Institute of Materials Research and Engineering, Agency for Science, Technology, and Research (A*STAR), 2 Fusionopolis Way, #08-03, Innovis, Singapore 138634, Singapore.

## ABSTRACT

Metasurfaces represent a pivotal advancement in nonlinear optics, leveraging high-Q resonant cavities to enhance harmonic generation. Multi-layer metasurfaces (MLMs) further amplify this potential by intensifying light-matter interactions within individual meta-atoms at the nanoscale. However, maximizing nonlinear efficiency demands extreme field confinement through optimized designs of large geometric and material parameters, which exceed traditional simulation's computational ability. To overcome this, we introduce NanoPhotoNet-NL, an AI-driven design tool employing a hybrid deep neural network (DNN) that synergizes convolutional neural networks (CNLs) and Long Short-Term Memory (LSTM) models. This framework accelerates nonlinear MLM design speed by four orders of magnitude while maintaining >98.3% prediction accuracy relative to physical simulators. The optimized MLMs achieve quality factors exceeding 50, enabling broadband third-harmonic generation (THG) in the deep ultraviolet (DUV) from wavelength 200 nm to 260 nm via parametric sweeps. Furthermore, NanoPhotoNet-NL facilitates dynamically tunable DUV nanolight sources with 20 nm spectral coverage in the UVC band using low loss nonlinear phase change materials. This work marks a transformative leap in nonlinear metasurface engineering, unlocking high-performance, reconfigurable platforms for nonlinear and quantum optical nanodevices.

* Corresponding author. Email address: Omar_Abdelrahman@imre.a-star.edu.sg

# 1. INTRODUCTION

The integration of photonics and nanotechnology fields have enabled huge performance enhancement in flat optics,[1-3] optoelectronics,[4-17] light modulation,[18-22] nonlinear optics,[23] and quantum optics devices.[24-26] Nonlinear optical metasurfaces have emerged as a transformative platform in nanophotonics, enabling ultrathin, compact, and highly efficient frequency conversion devices by tailoring light-matter interactions at subwavelength scales.[27] Among these, harmonic generation, particularly third-harmonic generation (THG), has garnered significant attention for applications in spectroscopy, deep ultraviolet (DUV) lithography, quantum optics, and bioimaging.[28-30] Recent progress in engineering high-Q resonances via bound states in the continuum (BICs), quasi-BICs, and multipolar interference has enabled metasurfaces to achieve enhanced field localization and nonlinear conversion efficiencies.[30] Despite these advancements, most designs are constrained to single-layer architectures, which inherently limit spatial confinement and modal overlap, thus capping achievable nonlinear efficiencies.

To circumvent these limitations, multilayer metasurfaces (MLMs) have been developed as a promising extension of planar optics, allowing vertical stacking of nanostructures to modulate both electric and magnetic modes across multiple layers.[31] This architectural complexity offers an expanded parameter space for achieving strong interlayer coupling, broader spectral tunability, and unprecedented control over nonlinear responses. Notably, MLMs have demonstrated enhanced Q-factors and field confinement, which is crucial for efficient nonlinear interactions at short wavelengths. Furthermore, the integration of tunable materials such as low-loss nonlinear phase change materials (PCMs) compounds introduces dynamic reconfigurability to MLMs, offering a new paradigm for on-demand tunable light sources in the ultraviolet domain.

However, designing MLMs for optimized broadband nonlinear response remains computationally prohibitive due to the vast combinatorial space of geometries, materials, and stacking configurations. Conventional numerical methods like finite-difference time-domain (FDTD) and rigorous coupled-wave analysis (RCWA), though accurate, scale poorly with the number of parameters, making exhaustive searches impractical. The challenge is further exacerbated when targeting broadband THG in the DUV (200–260 nm), where fine-tuning multiple geometric and material parameters is essential to maintain modal overlap and resonance enhancement across a wide spectral range.

To address this bottleneck, artificial intelligence (AI) and deep learning (DL) techniques have recently been introduced to expedite metasurface design.[32-34] Deep neural networks (DNNs) have been particularly effective in inverse design problems, enabling rapid prediction and optimization of nanophotonic structures without exhaustive simulations. For example, Peurifoy *et. al.* employed DNNs to model light scattering from silicon nanostructures,[35] while Liu *et. al.* used generative adversarial networks for on-demand photonic device synthesis.[36] In the nonlinear domain, Jing *et. al.* applied supervised learning to retrieve nonlinear susceptibilities of metasurfaces.[37] However, these approaches have predominantly focused on single-layer configurations, fixed materials, or narrowband responses, limiting their generalizability to complex MLMs architectures required for efficient broadband nonlinear conversion.

In this work, we present NanoPhotoNet-NL, a hybrid AI framework that leverages DNNs, convolutional neural networks (CNNs) and long short-term memory (LSTM) layers to model and optimize nonlinear MLMs for broadband DUV emission. By integrating CNNs to extract spatial correlations across geometric and material design parameters, and LSTMs to capture spectral dependencies in nonlinear response, our model enables accurate prediction of resonant modes across a wide wavelength range. The proposed framework accelerates the design process by four orders of magnitude compared to traditional simulators, with a prediction accuracy exceeding 98.3%. Using this platform, we demonstrate MLMs supporting Q-factors above 50 and producing tunable DUV THG emission from 200 to 260 nm. Moreover, by incorporating nonlinear PCMs antinomy trisulfide ($Sb_2S_3$),[38] we realize dynamically reconfigurable nanolight sources with 20 nm spectral tunability in the UVC band. This study marks a significant step toward AI-assisted design of advanced nonlinear photonic devices, paving the way for practical integration of compact, reconfigurable DUV sources in lithography, quantum optics, and biochemical sensing systems.

## 2. Methods

To enable robust optimization of nonlinear MLMs for THG DUV applications, we generated a comprehensive dataset spanning diverse geometric and material configurations. Each MLMs design comprised 1~5 vertically stacked layers with square-shaped nanopillars. Incorporated materials have distinct refractive indices including silicon dioxide ($SiO_2$), zinc oxide (ZnO), titanium dioxide ($TiO_2$), aluminum oxide ($Al_2O_3$), niobium pentoxide ($Nb_2O_5$), silicon nitride ($Si_3N_4$), amorphous antimony trisulfide (a-$Sb_2S_3$), crystalline antimony trisulfide (c-$Sb_2S_3$), and amorphous silicon (a-Si) as shown in Fig. 1a. Geometric parameters range such as layer heights ($h$) 20~100 nm, periods ($P$) 200~500 nm, and widths ($w$) 0.3*$P$~0.9*$P$ were uniformly sampled across their operational ranges. The optical response reflection spectra 500~800 nm was simulated using FDTD methods with $x$-polarized plane-wave excitation, periodic boundary conditions, and perfectly matched layers. Simulations employed a 10 nm mesh size, with field monitors positioned to capture reflection profiles. A total of 10,836 simulations were generated, divided into 70% as training data, 15% as validation data and as 15% test data.

Input data encoding leveraged MLMs symmetry as each 3D meta-atom was mapped to a 50×181-pixel grayscale image (10 nm/pixel resolution), where pixel intensity represented the refractive index of corresponding layers, substrate, or surrounding media as illustrated in Fig. 1b. Outputs consisted of reflection spectra discretized into 1000 points. All data underwent min-max normalization to [0, 1].

The NanoPhotoNet-NL framework synergistically combines DNN, CNN, and LSTM layers to efficiently model the spatial and spectral relationships inherent in MLMs designs. The CNN component operates as a spatial feature extractor through applying convolutional kernels to identify and learn hierarchical patterns in the geometric configuration of each metasurface layer. By incorporating convolutional and pooling operations, the model compresses the input dimensionality while retaining the essential spatial characteristics of the nanostructures, such as layer thickness and width. These encoded spatial features are then passed to the LSTM network, which excels at modelling temporal sequences. In this context, the LSTM captures spectral dependencies by processing the sequential optical response.

This hybrid architecture is especially suited to nanophotonic system design, where complex correlations exist between structure and optical performance across broad spectral domains. Table 1 outlines the specific architectural layers and corresponding tensor dimensions used in the model. The underlying convolution operation central to CNN is mathematically described as:

$$y_{i,j,k} = \sum_{m=0}^{K-1} \sum_{n=0}^{L-1} x_{i+m,j+n} \cdot w_{m,n,k} \qquad (1)$$

where *x* denotes the input tensor, *w* represents the learnable kernel weights, and *y* is the resulting feature map. Upon extraction of these features, the LSTM units model the dynamic behaviour of the optical response through its internal memory cell updates. The cell state ($c_t$) and hidden state ($h_t$) at time step *t* are computed using the following update rules:

$$c_t = f_t \odot c_{t-1} + i_t \odot \tilde{c}_t \qquad (2)$$

$$h_t = o_t \odot \tanh(c_t) \qquad (3)$$

In these expressions, $f_t$, $i_t$, and $o_t$ represent the forget, input, and output gates, respectively, while $\odot$ means element-wise multiplication. Together, these operations enable the model to retain long-range dependencies and learn wavelength-resolved optical behavior for enhancing its capability to predict and optimize nonlinear metasurface responses efficiently.

The framework was implemented in PyTorch with PyTorch Lightning for workflow optimization. Training used the Adam optimizer with learning rate = $10^{-3}$ and mean squared error (MSE) loss. Computations were performed on an NVIDIA GPU (8 GB VRAM) and 256 GB RAM workstation. Several model architecture tuning steps were applied to minimize MSE. Nonlinear harmonic generation physics were integrated inside NanoPhotoNet-NL to calculate the DUV THG for each designed MLMs. The spectral absorbed power ($P_{abs}(\omega)$) inside MLMs is calculated using volume integration of electric field intensity over MLMs unit cell and including imaginary permittivity ($\varepsilon''$) of each layer in equation 4. Third harmonic generation power ($P_{THG}$) in equation 5 is calculated using volume integration and third order susceptibility ($\chi^3_{(3)}$) of each material. Total DUV THG is calculated using equation 6 using literature measured values of following parameters interaction length ($l$), index at fundamental frequency ($n_\omega$), index at THG frequency ($n_{3\omega}$), and beam waist ($\omega_0$).

$$P_{abs}(\omega) = -\frac{1}{2}\omega \int_V \varepsilon''|E|^2 \qquad (4)$$

$$P_{THG}(3\omega) = 3\varepsilon_0 \int_V \chi^3_{(3)}(\bar{E}.\bar{E})\bar{E} \qquad (5)$$

$$P_{THG}(3\omega) = \frac{36l\chi^2_{(3)}}{\varepsilon_0^2 c^2 \lambda^2 n_\omega^3 n_{3\omega} \omega_0^4}\left[\int_\omega P_{abs}(\omega)\right]^3 \qquad (6)$$

**TABLE I**: NanoPhotoNet-NL model layers with total number of 2,740,816 trained parameters

| Layer type | Output shape | Parameters |
|---|---|---|
| Convolution (1 → 32) | [32, 177, 46, 20] | 832 |
| MaxPooling (2x2) | [32, 88, 23, 20] | 0 |
| Convolution (32 → 64) | [64, 84, 19, 20] | 51,264 |
| MaxPooling (2x2) | [64, 42, 9, 20] | 0 |
| Convolution (64 → 128) | [128, 38, 5, 20] | 204,928 |
| MaxPooling (2x2) | [128, 19, 2, 20] | 0 |
| Permute | [38, 128, 20] | 0 |
| LSTM (128 → 32) | [38, 32, 20] | 49,792 |
| Flatten | [1216, 20] | 0 |
| FCNN | [2000, 20] | 2,434,000 |

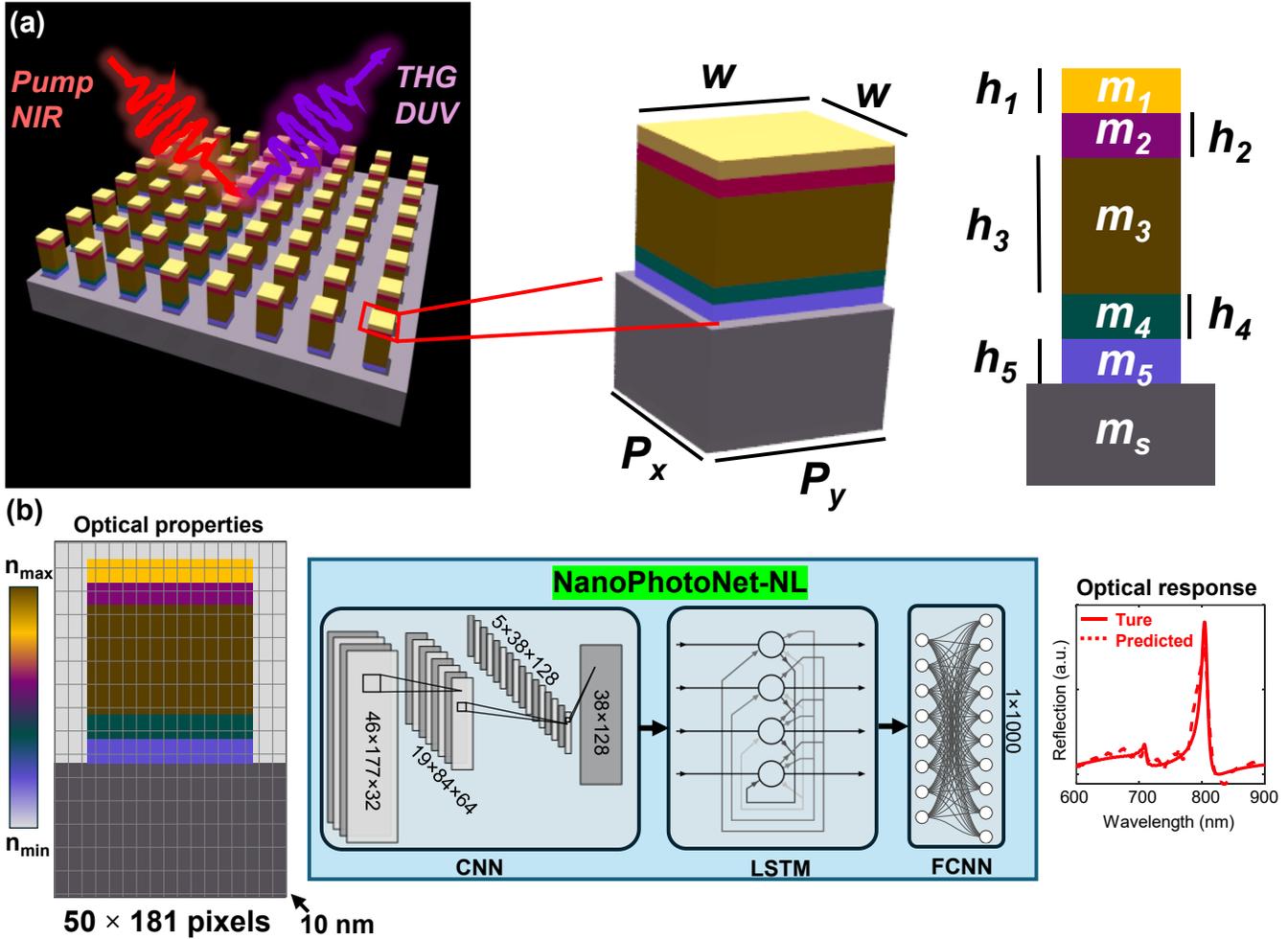

**FIG. 1.** Nonlinear MLMs using AI-assisted design approach. (a) Schematic of nonlinear MLMs with incident near infrared (NIR) pump laser and the reflected third harmonic generation in DUV. (Inset) a unit cell of nonlinear MLMs with symmetrical period ($P_x=P_y$) for simplifying design of 3D unit cell to 2D unit cell. Width ($w$), and each layer height named $h_i$ where $i$ correspond to number of layers and material $m_i$. (b) NanoPhotoNet-NL model architecture using grey-scale 2D image of size 50×181 pixels as input data. Layers of hybrid model consist of convolution, max pooling, flattened, LSTM, fully connected neural networks. Label data are reflection versus wavelength.

## 3. Results and Discussion

Figure 2 presents the evolution of the training and validation losses for the NanoPhotoNet-NL model across training epochs. The model exhibits a typical convergence behavior, where the training loss declines during the initial epochs as the model begins to capture underlying patterns in the data. In parallel, the validation loss also declines, albeit maintaining slightly higher values due to its role in evaluating model generalization on unseen samples. As training advances, the training loss plateaus at approximately 0.4% which indicates that the model has effectively minimized prediction error on the training set. Meanwhile, the validation loss continues to decline at a slower rate, eventually reaching a minimum of around 0.9%. The persistent yet narrow gap between training and validation loss demonstrates strong generalization capability, suggesting that the model avoids overfitting and performs reliably on previously unobserved MLMs configurations.

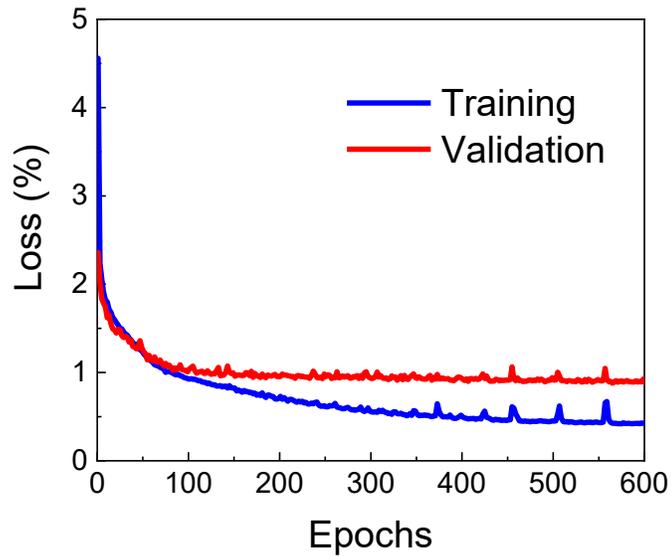

**FIG. 2.** The loss curve of NanoPhotoNet-NL model during training epochs.

**Section 3.1: Broadband DUV light emission**

To achieve broadband DUV emission via the parametric sweep approach, it is essential to design multiple MLM cavities that exhibit high-Q factors and strong field confinement within a nonlinear material at the fundamental wavelength range. The primary objective is to identify the optimal geometric parameters and material compositions that meet these performance requirements. Upon completing the model training, we employed NanoPhotoNet-NL to perform high-speed parametric optimization, selecting a-Si as the nonlinear material. This choice is motivated by its high third-order nonlinearity in the near-infrared (NIR) regime and its ability to support interband plasmon transitions in the DUV, which enhances THG through surface plasmon-induced field confinement.[30] The optimized unit cell structure, illustrated in Figure 3a, comprises five functional layers. The outermost layers are composed of $SiO_2$, forming a Fabry–Pérot-type cavity that promotes vertical field confinement within the pillar. The second and fourth layers consist of ZnO, serving as intermediate index layers to minimize abrupt refractive index transitions and suppress interfacial reflections. Centrally located is a 100 nm thick a-Si layer, strategically designed to absorb the incident NIR field and enhance nonlinear response. Both $SiO_2$ and ZnO layers have optimized thicknesses of 20 nm to achieve constructive interference and modal coupling. To generate multiple resonant modes within the fundamental wavelength range, the lattice period was swept from 230 nm to 380 nm while maintaining a constant width-to-period ratio of 0.5. The AI-predicted reflection spectra showed excellent agreement with FDTD simulations, achieving over 98.3% accuracy, as shown in Figure 3b. The resulting MLM structure demonstrated Q-factors up to 50 across a broad spectral window spanning 600 nm to 740 nm. The corresponding DUV THG response was calculated using Equation 5, and the spectral profile of the calculated harmonic signal is depicted in Figure 3c. A comparative analysis with an equivalent thin-film (TF) structure was constructed using identical layer materials and thicknesses but without nanoscale patterning has revealed a substantial enhancement in THG efficiency. Specifically, the optimized MLM achieved up to 500-fold improvement in DUV THG output relative to the unstructured TF counterpart. The DUV emission spans from 200 nm to 260 nm, and further wavelength tunability is attainable by tailoring the MLMs dimensions, enabling on-demand spectral engineering for specific photonic applications. Moreover, we benchmarked THG of our best design with THG in literature work using single layer $TiO_2$ metasurface. The literature achieved $Q$-factor of nearly 28 near 560 nm and our best MLMs have 77% enhancement in $Q$-factor, which shows the advantage of MLMs over single layer. Optimum MLMs could boost THG by more than five-fold enhancement.[39]

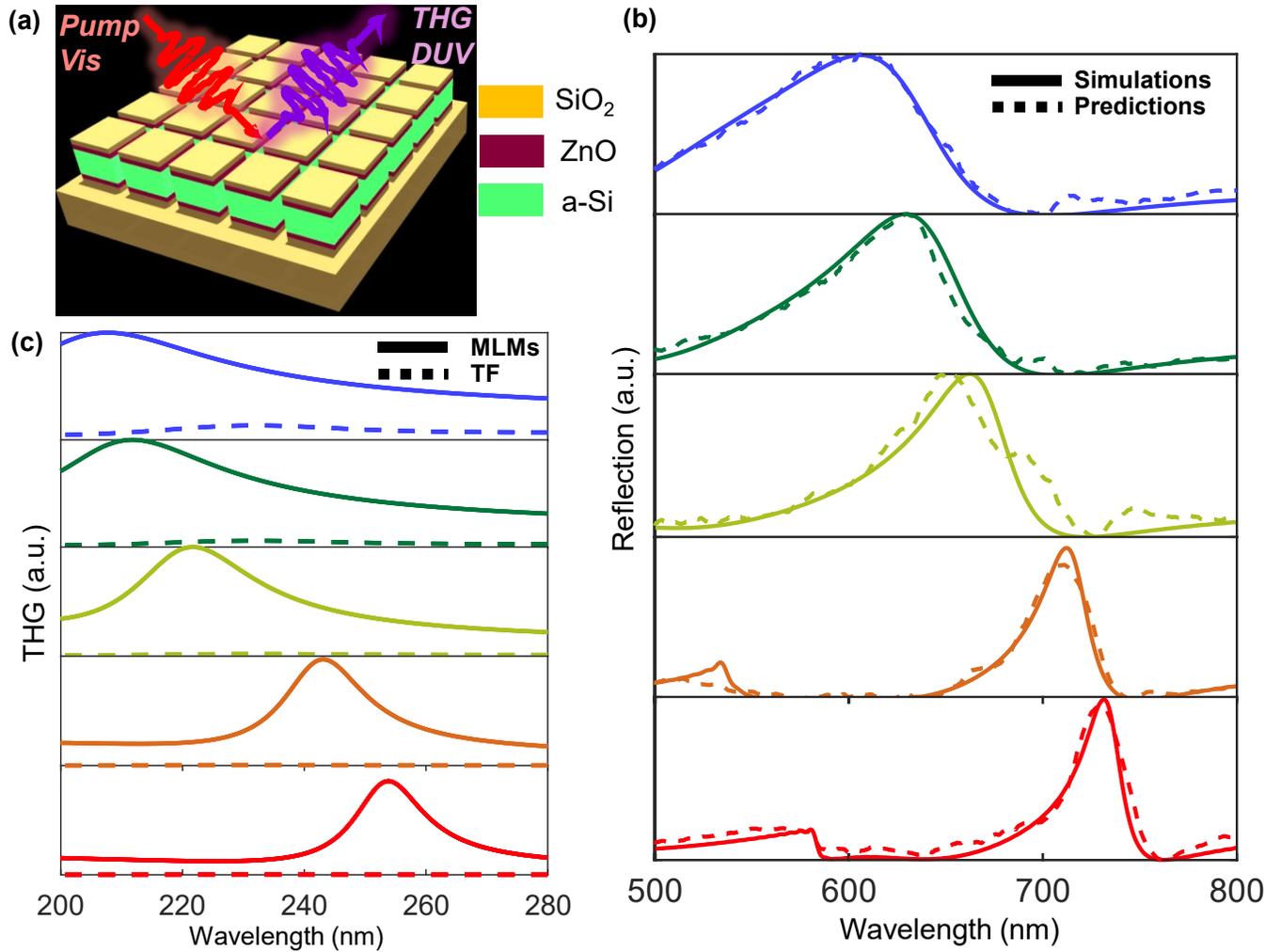

**FIG. 3.** Broadband DUV light emission in nonlinear MLMs through parametric sweep. (a) 3D schematic of nonlinear MLMs with intermediate a-Si layer between $SiO_2$ and ZnO layers for strong light confinement in a-Si through Fabry-Perot resonance. (b) Simulated and predicted reflection spectrums for nonlinear MLMs using different periods for broadband generation of DUV light. (c) Simulated THG emission of the corresponding MLMs in Fig. 3b compared with thin film (TF) THG.

**Section 3.2: Ultra-fast reconfigurable DUV nano-light sources**

While broadband DUV emission through parametric sweeping offers flexibility, it is not ideal for all applications specially when compactness and dynamic control are prioritized. For future integrated flat optics and miniaturized devices, there is a strong demand for actively tunable DUV emission from a single metasurface cavity. Rapid wavelength tunability would be especially beneficial in high-resolution, time-resolved photoluminescence imaging, enabling advanced functionality in bioimaging and sensing. One promising approach to meet these requirements involves incorporating nonlinear $Sb_2S_3$, which combines low optical losses in the visible range with strong third-order nonlinearity, making it well-suited for efficient THG in the DUV regime. Additionally, $Sb_2S_3$ exhibits a substantial refractive index contrast between its amorphous and crystalline states ($\Delta n > 1.2$), providing a wide tuning window for optical resonances. Importantly, phase transitions in $Sb_2S_3$ can be triggered with a single picosecond laser pulse (~10 ps) via melt-quenching, enabling switching speeds up to 100 GHz, as shown in Figure 4a.[20] Leveraging the trained NanoPhotoNet-NL model, we performed rapid parametric optimization using $Sb_2S_3$ as the nonlinear core material. The resulting reflection spectra for MLMs with lattice periods ranging from 200 nm to 400 nm are shown in Figure 4b. The predicted resonance features exhibit excellent agreement with full-wave FDTD simulations, confirming the model's reliability. The corresponding THG enhancement, plotted in Figure 4c, demonstrates up to a 790-fold improvement compared to unpatterned multilayer $Sb_2S_3$ thin films with identical thicknesses. Upon inducing a phase switching from the amorphous to crystalline state, the refractive index of $Sb_2S_3$ increases, leading to a noticeable redshift in the resonance wavelength for each device configuration, as depicted in Figure 4d. Although the transition introduces a slight broadening in the reflection peaks because of the increased optical losses in the crystalline phase, the spectral tunability remains robust. As shown in Figure 4e, the THG has up to 470-fold enhancement over thin film counterpart and the output THG can be dynamically tuned across a 20 nm range. Demonstrated results enable reconfigurable and on-demand DUV light generation within a single and compact nonlinear MLMs.

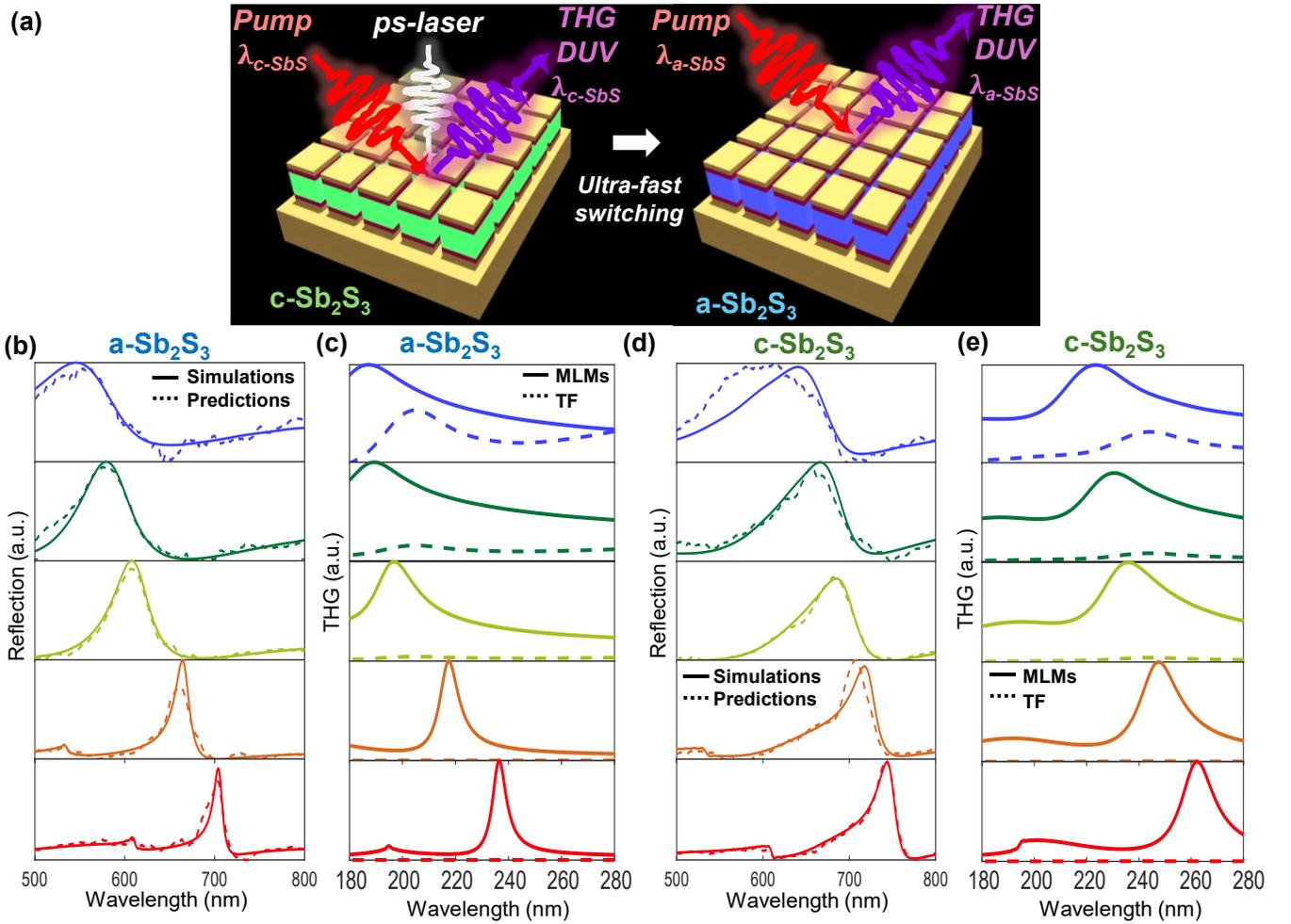

**FIG. 4.** Tunable DUV light emission in nonlinear MLMs through phase change material. (a) 3D schematic of ultrafast reconfigurable nonlinear MLMs using $Sb_2S_3$ between ZnO and $SiO_2$ layers to enhance light confinement inside individual pillar. Pulsed picosecond laser (ps-laser) switches the phase of $Sb_2S_3$ ultra-fastly from crystalline (c-$Sb_2S_3$) to amorphous (a-$Sb_2S_3$). (b) The left column shows the predicted and the simulated reflection of nonlinear MLMs at a-$Sb_2S_3$. The right column plots the calculated THG of nonlinear MLMs at a-$Sb_2S_3$ compared with thin film structure using same thickness. (c) The left column illustrates the corresponding predicted and simulated reflection spectrum after the phase changes to c-$Sb_2S_3$ with more than 60 nm tuning range in NIR reflection. The right column plots the calculated enhanced THG of nonlinear MLMs at c-$Sb_2S_3$ compared to TF with large tunable DUV THG emission range above 20 nm.

Following the spectral tuning analysis of THG in $Sb_2S_3$-based MLMs, we computed the total generated DUV THG power using Equation 6. To ensure a meaningful comparison, we calibrated the THG output from TF $Sb_2S_3$ structures against literature.[40] The results reveal that the THG power from a-$Sb_2S_3$ is approximately seven times lower than that of c-$Sb_2S_3$, owing to the significantly weaker third-order nonlinear susceptibility of the amorphous phase. Specifically, under a 1 mW pump, the THG output from TF a-$Sb_2S_3$ was found to be approximately 0.4 pW, whereas TF c-$Sb_2S_3$ generated ~2.8 pW. In contrast, the MLMs configurations led to a dramatic increase in THG power, a-$Sb_2S_3$-based MLMs yielded up to 1.02 nW, and c-$Sb_2S_3$-based MLMs produced as much as 400 nW, as shown in Figure 5a. The enhancement by more than two orders of magnitude compared to thin-film structures underscores the efficacy of MLMs in boosting nonlinear light generation. Such high THG output paves the way for compact, high-performance DUV nanolight sources suitable for biomedical imaging and advanced lithography systems. Moreover, multilayer metasurfaces offer a significant advantage over traditional single-layer metasurfaces due to stronger optical field confinement and enhanced nonlinear interactions. In fact, THG generation in MLMs scales nonlinearly with increased absorption of the fundamental field, enabling up to three orders of magnitude improvement in THG output. For instance, single-layer $TiO_2$ metasurfaces typically yield THG power on the order of 100 pW at 1 mW pump power as illustrated in Figure 5b.[39] In comparison, our nonlinear MLMs demonstrate at least a tenfold increase in THG output, reinforcing their superior performance for next-generation nonlinear nanophotonic platforms. Lastly, we compared our proposed AI-model prediction efficiency with literature papers using DNN-based AI models and summarized them in Table 2. Our model have better accuracy than literature models.

**Table II:** Performance comparison between literature AI models and NanoPhotoNet-NL model.

| AI architecture | Accuracy (%) |
|---|---|
| DNN [41] | 86 |
| NanoPhotoNet-NL | 98.3 |
| DNN [42] | 92.4 |

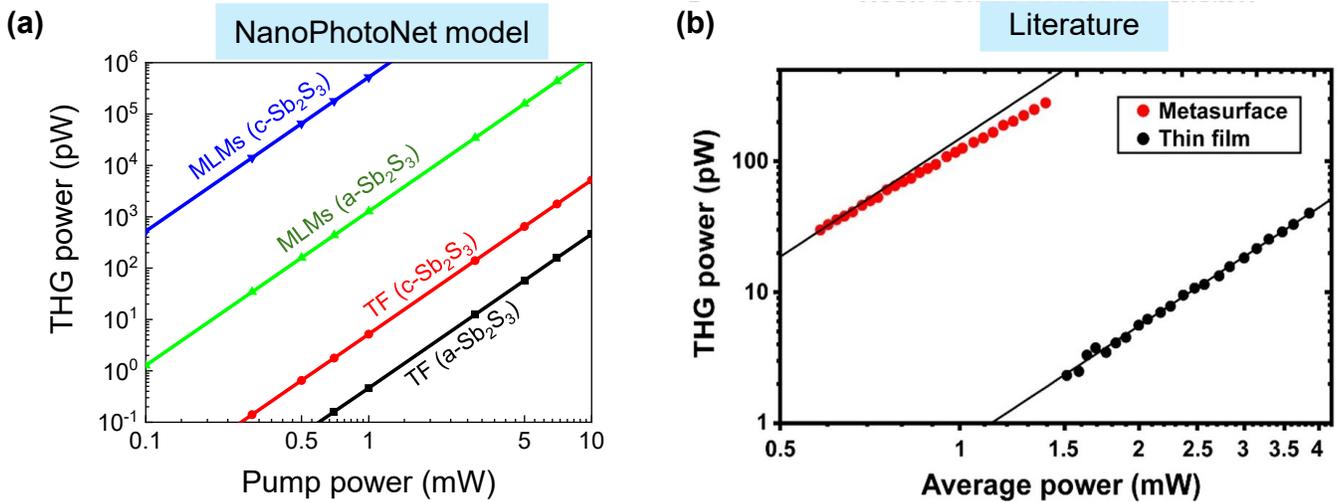

FIG. 5. THG power of nonlinear MLMs versus nonlinear metasurface in literature.[39] (a) Calculated THG power vs pump power for optimized nonlinear MLMs at amorphous and crystalline states compared with THG of their corresponding thin films enabling THG of more than 1 nW at pump power of 1 mW. (b) THG of TiO$_2$ metasurface compared with their thin film showing THG of 100 pW at pump 1 mW. Reprinted with permission from American Chemical Society, Copyright 2019.

## 4. Conclusion

In conclusion, we have developed NanoPhotoNet-NL, an advanced AI-powered framework that significantly accelerates and enhances the design of nonlinear MLMs for DUV-THG. By integrating DNNs, CNNs, and LSTM architectures, NanoPhotoNet-NL overcomes the limitations of conventional simulation-based optimization. NanoPhotoNet-NL allows a speed-up of four orders of magnitude with a prediction accuracy exceeding 98.3%. The designed nonlinear MLMs exhibit quality factors above 50, enabling efficient and broadband THG from 200 to 260 nm. Moreover, the inclusion of low-loss nonlinear PCMs allows for dynamically tunable emission with 20 nm spectral coverage in the UVC band. These capabilities establish NanoPhotoNet-NL as a transformative tool for advancing high-performance, reconfigurable nonlinear nanophotonic devices. Looking forward, the integration of physical constraints from nonlinear and quantum optics into the AI model architecture could enable the co-optimization of quantum light generation processes. This would open new avenues for AI-assisted discovery of entangled photon-pair sources, squeezed light emitters, and nonlinear quantum metasurfaces, paving the way for next-generation technologies in quantum communication and quantum computing.

**Credit Authorship Contribution Statement**

**Omar A. M. Abdelraouf:** Project administration, Methodology, Conceptualization, Investigation, Data curation, Validation, Supervision, Resources, Visualization, Writing – original draft & review.

**Declaration of Competing Interest**

The authors declare no competing financial interests or personal competing interest.

**Data Availability**

NanoPhotoNet-NL code and raw data are available on request

**Acknowledgments**

Author acknowledges the Agency for Science, Technology, and Research (A*STAR) for the scholarship provided Singapore International Graduate Award (SINGA).